# Long-range, phase-and-polarization diversity coherent reflectometer

P. CHO,[1] Y. LENG,[1,*] , P. PETRUZZI,[2] M. MORRIS,[3] G. BAUMGARTNER,[3] AND J. GOLDHAR[1]

[1]*Electrical and Computer Engineering Department, University of Maryland, College Park, MD 20742, USA*
[2]*Laboratory for Physical Sciences,8050 Greenmead Drive, College Park, MD 20740, USA*
[3] *Laboratory for Telecommunication Sciences,8080 Greenmead Drive, College Park, MD 20740, USA*
**yleng@umd.edu*

**Abstract:** A coherent reflectometer with a short-length reference arm, which utilizes a 76-MHz repetition rate mode-locked fiber laser, was investigated experimentally for long fiber links (> 10 km). The reflectometer combines advantages of optical time-domain and frequency-domain reflectometry without the need for high-speed photodetectors and electronics to achieve high spatial resolution of 2.5 mm at 10 km. To our knowledge, this is the highest resolution reported for fiber reflectometry at this distance for reflective type event using non-photon-counting detection. Phase and polarization diversity detection, combined with spectral compression using frequency-chirp, are proposed to improve sensitivity and discrimination against the fiber Rayleigh backscattering background.

## 1. Introduction

In conventional optical time-domain reflectometry (OTDR) for long fiber links (> 10 km) using short laser pulses, the spatial resolution is limited by the launched pulse duration and the fiber dispersion. For example, a commercially-available OTDR has a spatial resolution in the range of meters. Achieving this resolution requires fast detectors and electronics to resolve arrival time differences between reflected pulses [1]. An OTDR using photon-counting detection can achieve centimeter resolution over test distances of tens of kilometers [2].

Optical frequency-domain reflectometry (OFDR), on the other hand, does not require high-speed detection to achieve high spatial resolution, but its testing distance is limited. For example, one commercially available reflectometer is capable of operating over a fiber length of up to 2 km with the best resolution of 0.25 mm [3].

Another approach to reflectometry, optical low coherence reflectometry (OLCR), achieves spatial resolution less than 10 mm using a reference arm with an adjustable optical delay line [1]. For OLCR to function over long test fiber lengths, the reference arm length must match the test arm length within the coherence length of the light source. Optical coherence tomography (OCT), widely used in medical imaging applications, is a form of short-range but high-resolution reflectometry that employs techniques of OFDR and/or OLCR.

In this paper, we describe an interferometric reflectometer with phase and polarization diversity detection, combining elements of OTDR and OFDR. We use it to demonstrate spatial resolution less than 2mm over a 10km fiber length. This reflectometer uses optical pulses, but does not require high-speed photodetectors or receiver electronics. Unlike traditional OTDR and OFDR, the required bandwidth of the reflectometer detector is independent of spatial resolution and test distance. And unlike OLCR, our approach does not require length matching the of test and reference arms. This last feature was achieved by fine tuning the pulse repetition rate (PRR) of a mode locked laser to match the arrival times of the returned test pulses and the reference pulses at the interferometer output, as described in detail later.

For long fiber lengths the ultimate resolution is limited by the environmental perturbations on the fiber link optical path length during the measurement time. For discrete reflective events, using dispersion compensation in a short-length (few meters long) reference arm, we observed resolutions of 2.5 mm at 10 km and 1 cm at 50 km distances. Resolution as defined here is the minimum resolvable distance between two discrete reflective targets. To our knowledge, this is the highest resolution reported for fiber reflectometry at this distance using non-photon-counting detection. The proposed reflectometer complements OTDR and OFDR type instruments and fills application gaps that require long-range high-resolution reflection measurements.

Matching the dispersions of the test and reference arms is crucial to achieving the observed spatial resolution. This matching can be done using a commercial off-the-shelf telecom tunable dispersion compensator (TeraXion TDCMB). The spatial resolution is limited by the input pulse spectral width $\Delta\lambda$, and is independent of the pulse temporal width. However, as will be described later, there are practical factors that limit the spatial resolution beyond what is predicted by the spectral width.

With shot-noise-limited detection, the proposed reflectometer's sensitivity, defined as the minimum detectable target reflectivity, is primarily limited by Rayleigh-backscattering (RBS) in the fiber. This limitation is a result of the laser pulse repetition interval being much shorter than the round-trip time to the end of the fiber and creates a constant background signal that acts as the noise floor of the system. The detected signal is therefore composed of the desired target-induced reflection as well as the incoherent sum of the distributed RBS light along the fiber created by multiple laser pulses. The noise background that results from detecting RBS from a large number of pulses can be several orders of magnitude greater in our reflectometer than in conventional reflectometers.

In this paper we will demonstrate two different techniques for reducing the RBS background and improve the sensitivity. One uses Stokes' vector averaging and the other laser phase compensation, similar to the Frequency Modulated Continuous Wave (FMCW) technique used in RF radar [4].

## 2. Coherent detection with high repetition rate pulses

Fig. 1 shows the basic schematic for the proposed coherent reflectometer. A commercial passively mode-locked fiber laser (MLL) from PriTel, Inc. was used as the pulse source with a nominal PRR of 76 MHz which results in a pulse separation of 13.2 ns, that is orders of magnitude shorter the round trip time in the fiber link test arm. The PRR can be tuned over a range of about 174 kHz by adjusting the laser cavity length using a micrometer. A tunable band-pass optical filter (BPF), not shown in the figure, is used to limit the pulse spectral width to about 0.5 nm.

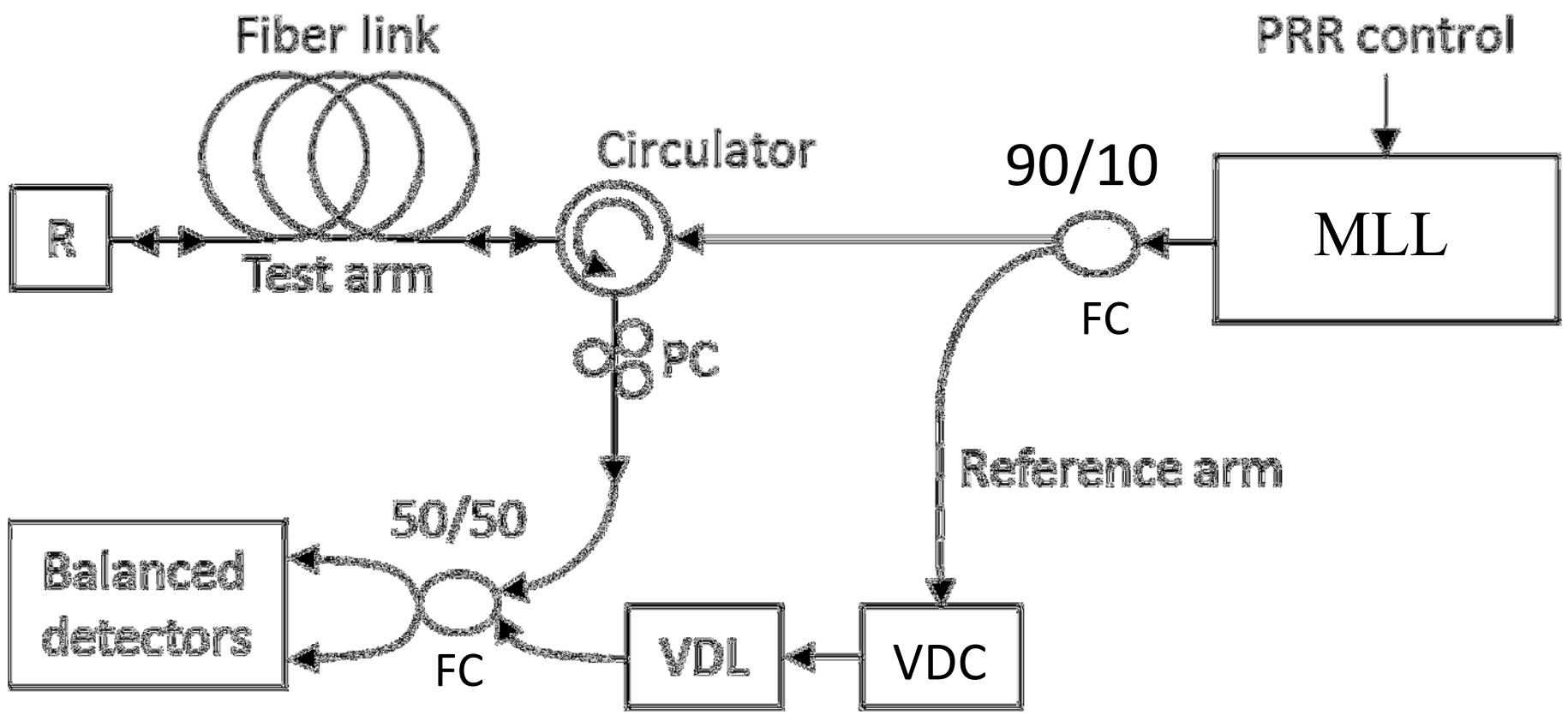

Fig. 1. Simplified schematic of our reflectometer. The input laser pulse source on the far right is followed by an interferometer with a directional coupler with a power splitting ratio of 10:1. The reflective target event at the end of the fiber link is denoted by R on the far left. MLL:mode locked laser. VDL: variable optical delay line. VDC: variable dispersion compensator. PC: polarization controller. PRR: pulse repetition rate. FC: fiber coupler.

As shown in Fig. 1, the majority of the output of the MLL is used to probe the target in the fiber link test arm using a 90/10 fiber coupler. The remaining portion of the laser output is used as a local oscillator in the reference arm, which has a variable dispersion compensator and a variable delay line. The returned signal from the fiber link test arm is combined with the local oscillator using a 50/50 fiber coupler and then detected with a balanced detector. The electrical output of the balanced detectors is the returned target signal time-gated by the local oscillator pulses. Only the points corresponding to the peaks of the local oscillator pulses are selected and then numerically Fourier transformed. Processing the reflectometer signal in this manner results in the magnitude and shape of the spectrum changing significantly when the PRR or the variable delay line in the reference arm is tuned to detect a target.

The spectrum, measured in the manner described above, for a reflection from a flat cleaved end of a 10 km long of SMF-28 fiber spool is shown by the red trace in Fig. 2. The broader spectral width of the red trace is the signal observed when the pulse repetition rate and variable delay line are set so the reflected and the local oscillator pulses overlap in time at the balanced detectors. The blue trace is the spectrum observed when either the fiber is terminated with an angle-polished connector or when the pulses do not overlap at the detector. This trace shows the Rayleigh backscattering (RBS) from the target position as well as from the approximately $7\times10^3$ laser pulses which are separated by 13.2 ns. The spectral width obtained varies linearly with the length of the tested fiber. This spectral broadening is due to laser phase variations and fluctuations in the fiber's optical path length due to thermal and mechanical perturbations. Due to the distributed nature of Rayleigh backscattering, the entire fiber length, regardless of where the target is located, contributes to the RBS background noise. The black trace in Fig. 2 is the receiver noise signal obtained when the test fiber link is removed from the reflectometer, and is primarily due to detector shot noise and electronic thermal noise. The noise spike in the spectrum at DC is due to a very small unbalance in the receiver detector pair.

The ability of this reflectometer to detect weakly reflecting targets is clearly impaired by the enhanced RBS noise, which cannot be suppressed by improving the detector sensitivity or by using a long integration time. Two different techniques for mitigating the RBS noise were developed. One technique involves Stokes vector averaging of the polarization-diverse and phase-diverse reflectometer signal, and the second technique utilizes an auxiliary interferometer to monitor the laser phase noise. The laser phase variations measured in the auxiliary interferometer can be used to correct the fluctuations of the signal reflected from a single target, leading to spectral compression after the FFT. The Rayleigh scattering background spectrum, on the other hand, remains broad and does not get compressed after this correction.

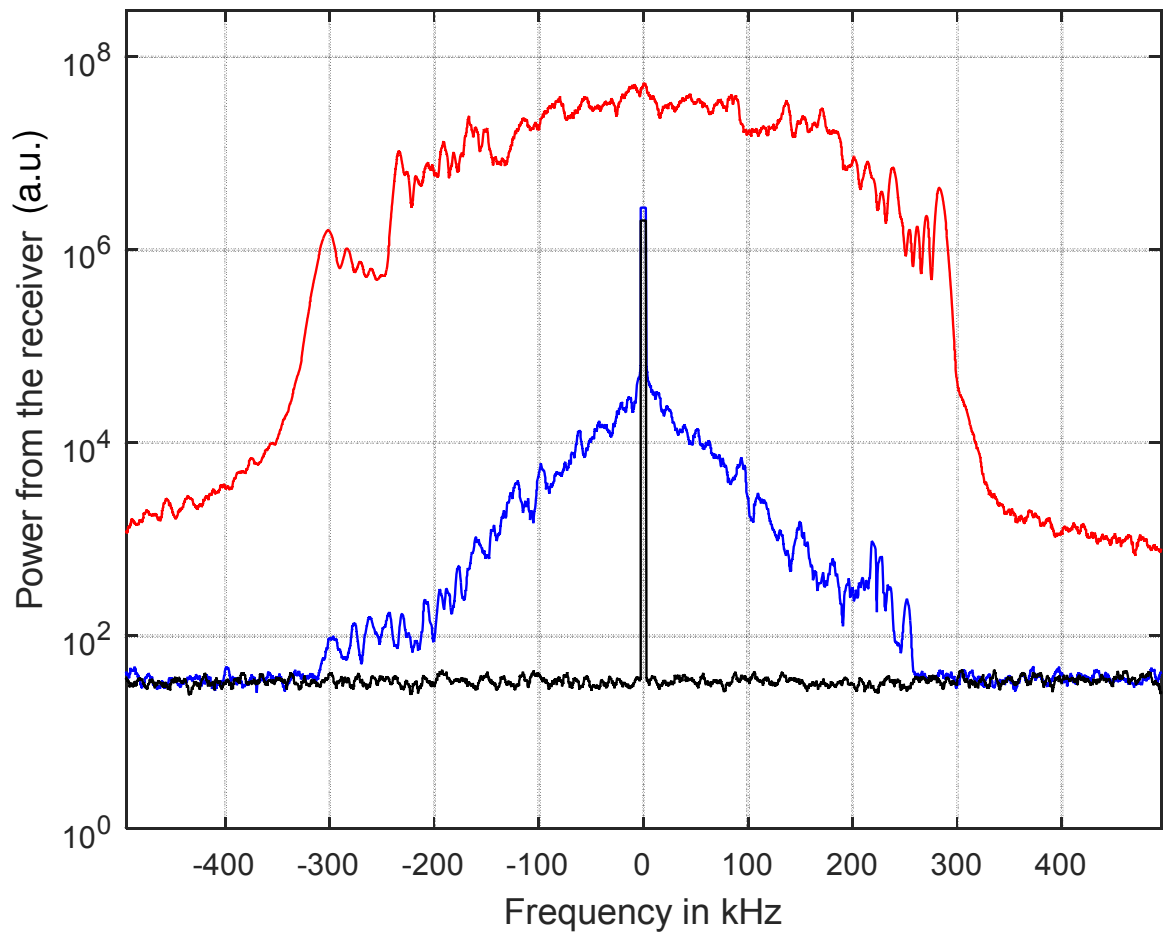

Fig. 2. RF power spectra of the coherent detector output observed with a target at the end of a 10 km fiber. The top curve (red) is with the target, the middle curve (blue) is fiber without the target, and the bottom curve (black) is without any fiber. The spike in the center of the black curve is due to a small imbalance between the two detectors in the receiver. Each data set contains approximately 20 msec of signal samples. Each data point in the plot is a spectral average over 5kHz.

The ultimate sensitivity of a coherent reflectometer is determined by calculating the signal to noise ratio over a time interval, $T_{coh}$, for which the system is stable. When the laser frequency is stable, fluctuations in the test arm fiber length determine the coherence time interval $T_{coh}$. When the fiber length is stationary but the laser frequency fluctuates, $T_{coh}$ will be determined by the rate of laser phase fluctuations. Within the time period $T_{coh}$, reflectometer measurements are made using coherent detection with a local oscillator signal strong enough to achieve shot noise limited detection (thermal and Laser Relative Intensity Noise (RIN) noise are ignored). The shot-noise limited Signal-to-Noise Ratio is then approximately given by the number of signal photons detected during $T_{coh}$. For a group of $N$ samples, where each sample is measured over a time period of $T_{coh}$, but each sample is incoherent with the other samples, the total SNR will be $SNR \approx \sqrt{N} P_s \cdot T_{coh}/\hbar\omega$.

The coherence time, estimated from our data to be on the order of $T_{coh} \approx 100$ µs, is usually limited by small fluctuations in the length of the fiber, which leads to changes in the phase of the signal. However, for our reflectometer a much larger source of phase variations between the signal and the local oscillator is the frequency variation of the MLL source. As will be described in more details in Section 4, the laser phase noise can be monitored and compensated for using signal processing.

Based on the above discussion, higher laser power would result in higher detection sensitivity, or allow a shorter acquisition time for the same sensitivity. However, optical nonlinearity in the fiber limits the amount of laser power that can be launched and still maintain the linear propagation required to achieve optimal resolution. When the launched power level into the fiber link test arm is below ~ −10 dBm, fiber nonlinearities can be neglected.

As can be estimated from the data in Figure 2, fiber RBS limits the detectable target reflectivity at 10 km to $r \sim 10^{-6}$ in our system without significant signal processing. By taking advantage of the laser's phase modulation this can be improved by at least two orders of magnitude. Thus, our reflectometer, with a large number of data samples, should be able to observe at 10 km a target reflectivity as low as $r \sim 10^{-8}$. Additional improvements are possible with increased laser frequency chirp and by taking advantage of the fact that the polarization of RBS from multiple points along the signal path will vary on a time scale longer than $T_{coh}$.

The ultimate spatial resolution of our reflectometer is determined by the laser source's spectral width, $\Delta\lambda$, centered around $\lambda$. This is given by [1]

$$\Delta z_r \cong \frac{1}{2n_g}\frac{\lambda^2}{\Delta\lambda},$$

where $n_g$ is the fiber group index of refraction. With pulses of bandwidth ~0.5 nm, a resolution of ~$(2.405/n_g)$ mm for $\lambda$ = 1.55 μm is expected, and this resolution is in fact observed in short fiber lengths. However, the experiment requires long measurement times, on the order of seconds or even minutes. During that time, for long fibers, thermal drifts in the test fiber can be significant and result in less precise measurement of position.

It is impractical to use a coherent reflectometer to measure targets along the entire length of a long fiber link test arm because of the requirement to have equal test and reference arm lengths in the interferometer. Our interferometer, however, can have very different arm lengths by taking advantage of the equal spacing between the pulses from the MLL. This allows interference between the returning reflected signal pulses and the local oscillator pulses that were emitted at a much later time by the MLL.

One advantage of implementing the interferometer with unequal arm lengths is the ability to search for target reflections by scanning the fiber through large incremental distances using only slight changes in the pulse repletion rate. The PRR of the laser is adjusted to ensure maximum temporal overlap of the reflected light from the target and reference pulses at the interferometer output, which is similar to the Vernier effect. Since this a periodic effect, one can calculate the maximum PRR variation or tuning range required to achieve one period, which allows for the detection of an arbitrarily positioned target in the fiber link test arm. When the length of the fiber to the target is equal to $\Delta L$, changing the time delay between laser pulses ($T_l = (76\,MHz)^{-1} \approx 13.2\,ns$) by a small amount, $\delta t$, moves the observation point in time by $\delta t \times T_{RT}/T_l$, where $T_{RT} = 2\Delta L/v_g$ is the round trip propagation time and $v_g$ is the propagation group velocity in the fiber.

For the full tuning range we have $\delta t \times T_{RT}/T_l = T_l$, and the length of the laser cavity has to be changed by about $\delta l = v_g\,\delta t \approx 2$ mm for $\Delta L = 10$ km. This change can be readily accomplished with a mechanical variable optical delay line. For longer distances to the target, the full tuning range is achieved with smaller laser cavity length changes.

Once the reflective target is detected by tuning the PRR, we can perform a scan of spatial reflectivity variations along the fiber by tuning the Variable Delay Line (VDL), which can change by about 10 cm. In this measurement, there is an ambiguity in the absolute distance to the target because, at the peak of the reflectivity, $\Delta L$ is an integer multiple of laser cavity lengths. The ambiguity can be easily removed, and the absolute distance to target can be determined, by changing the delay in the VDL and finding a new PRR.

## 3. Phase and polarization diversity reflectometer

Fig. 3 shows a more detailed schematic of our experimental setup. A fast four-channel digitizing oscilloscope (Tektronics DPO5104) was used to record the I and Q output channels of a coherent receiver, the input polarization monitor, and the output of an auxiliary Mach-Zehnder interferometer typically at sampling rate of 5 or 10 GSample/sec.

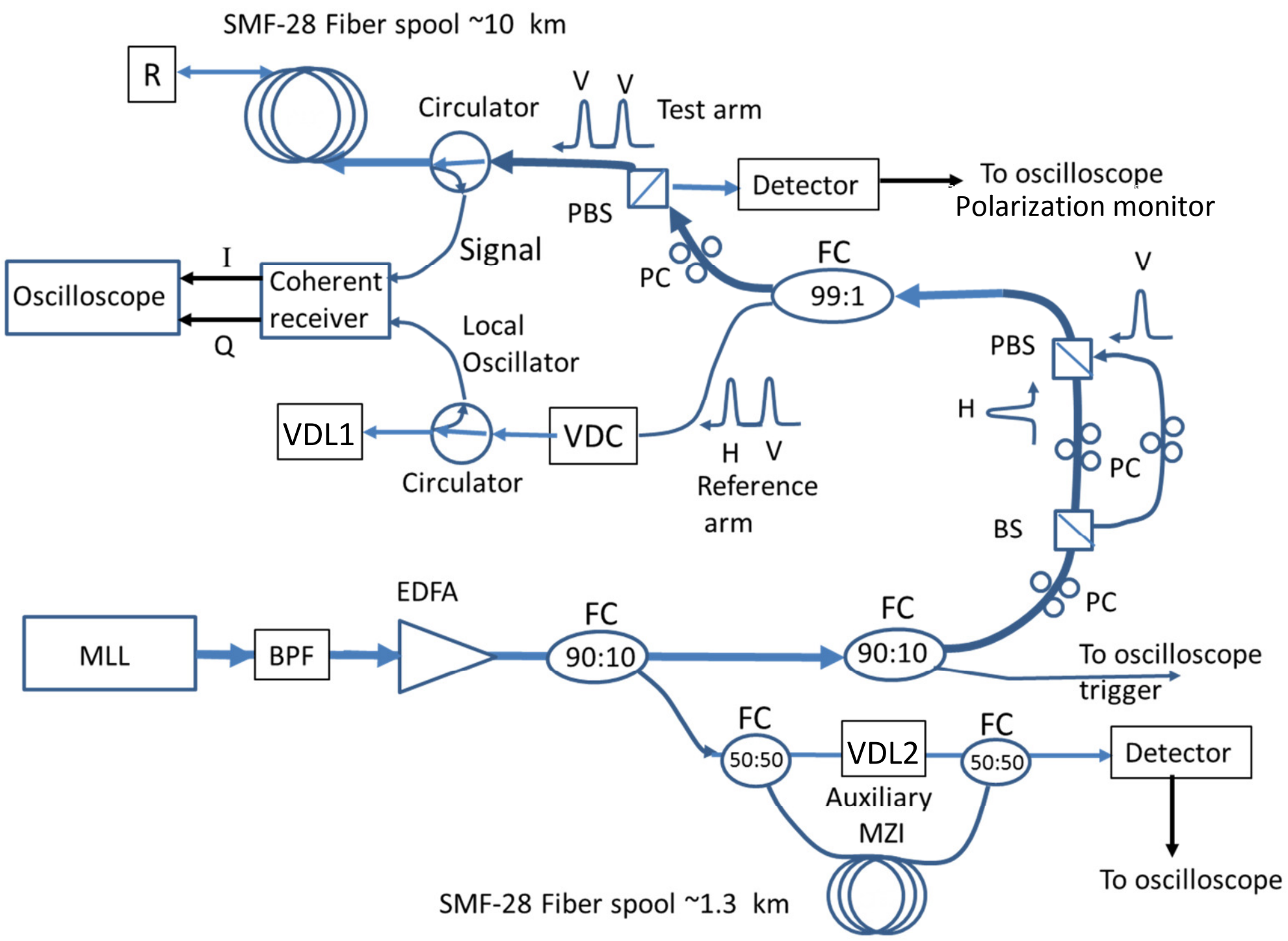

Fig. 3. Schematic of phase and polarization diversity reflectometer. BS: beam splitter, PBS: polarization beam splitter, PC: polarization controller, VDL1/2: variable optical delay lines, VDC: variable dispersion compensator, R: reflective target MLL: mode locked laser, BPF: band pass filter, EDFA: Erbium doped fiber amplifier, MZI: Mach-Zehnder Interferometer, FC: fiber coupler.

The pulse shape and pulse duration were not measured, but the duration is estimated to be about 5 ps based on the 0.5 nm spectral width, defined by the band-pass filter, and an assumed un-chirped $sech^2$ pulse shape, yielding a duty cycle of 0.038%.

The phase variation of the laser pulses was monitored using an auxiliary Mach-Zehnder Interferometer (MZI) with a ~1.3 km path length difference. A PIN diode detector was used to measure one output of the interferometer. Signal processing of the MZI output allowed the phase variation of the mode locked laser pulses to be calculated as a function of time.

Polarization diversity in the reflectometer was implemented by time-interleaving orthogonally-polarized laser pulses in the reference arm with a known preset delay, $\tau_{VH}$, between the pulses, as shown in Fig. 3. This approach avoids the need for a second coherent receiver, and thereby reduces the complexity of the reflectometer. Separation of the two polarizations can be easily performed numerically on the digitized signals since $\tau_{VH}$ is known. In order to ensure proper polarization diversity detection, the time-multiplexed pulses in the test arm all have the same polarization.

The reflectometer uses a Michelson fiber interferometer with test and reference arms as depicted in Fig. 3. Standard single-mode fiber (Corning SMF-28e+) with lengths up to 50 km were used in the test arm, followed by a reflective target which was composed of two reflectors, both with ~ 1% reflectivity, separated by ~ 8 mm. The two-reflector target yields a unique double-peak signature in the delay scan, which allows unambiguous location determination. At the reference arm, a computer-controlled variable optical delay line (VDL1) with a 1120 ps double pass delay range was used to align in time the reflected optical signal and the local oscillator pulses.

To match the test fiber dispersion, a commercial tunable dispersion compensator (TDC) from TeraXion (benchtop tunable dispersion compensating module, TDCMB) is inserted in the reference arm. The TDC employs a pair of matched chirp fiber Bragg gratings on a thermal gradient platform and is channelized for Dense Wavelength Division Multiplexing (DWDM) telecom applications with a channel bandwidth of about 0.5 nm. As a result, in order to minimize the impact on spatial resolution, the spectral width of the laser source is limited to about 0.5 nm using a band-pass filter. The adjustable dispersion range is ±900 ps/nm and the effective or useable channel width decreases linearly from 0.6 to about 0.4 nm. This is one of the factors that contributes to the degraded resolution observed for long fiber lengths which require large dispersion matching.

Phase diversity detection was achieved by using a coherent receiver (OptoPlex, Inc.), which is composed of an optical 90° hybrid followed by a pair of balanced photo-receivers that output an electrical signal consisting of the in-phase (I) and quadrature (Q) components of the input optical signal. The bandwidth of the balanced photo-receiver is 100 MHz, with a nominal conversion gain of 2250 V/W (2120 to 2367 V/W) and a nominal DC Common Mode Rejection Ratio (CMRR) of 20 dB (24 to 27 dB). The worst case phase error is a -1.829° deviation from 90°.

The digitized I and Q signals from the oscilloscope at a sample rate of 5 GSamples/s were down sampled at the 76 MHz pulse rate. And a complex reflected signal $I(t)+iQ(t)$ was synthesized from two real $I(t)$ and $Q(t)$ signals. To accurately down sample the acquired signals, the precise delay between the trigger pulse and the I and Q signals was determined through cross-correlation of the reflected signal and the trigger pulse captured simultaneously at the output of the input polarization monitor. The time lag corresponding to the peak of the cross-correlation curve yields the delay, which is used to align the complex reflected signal to the laser pulse. The time locations of the all the laser pulse peaks are then extracted from this signal and the complex $IQ$ signal was then down sampled at these peak time locations.

Fig. 4a shows a snap-shot of a typical complex reflected signal amplitude, $|I(t) + iQ(t)|$, before down sampling and $IQ$ plots for the two orthogonal polarizations after down sampling and being separated numerically. As can be seen in Fig. 4b, the imperfections of the coherent receiver (i.e., the phase difference deviates from 90° and it has a finite CMRR) yields off-centered non-circular $IQ$ orbits.

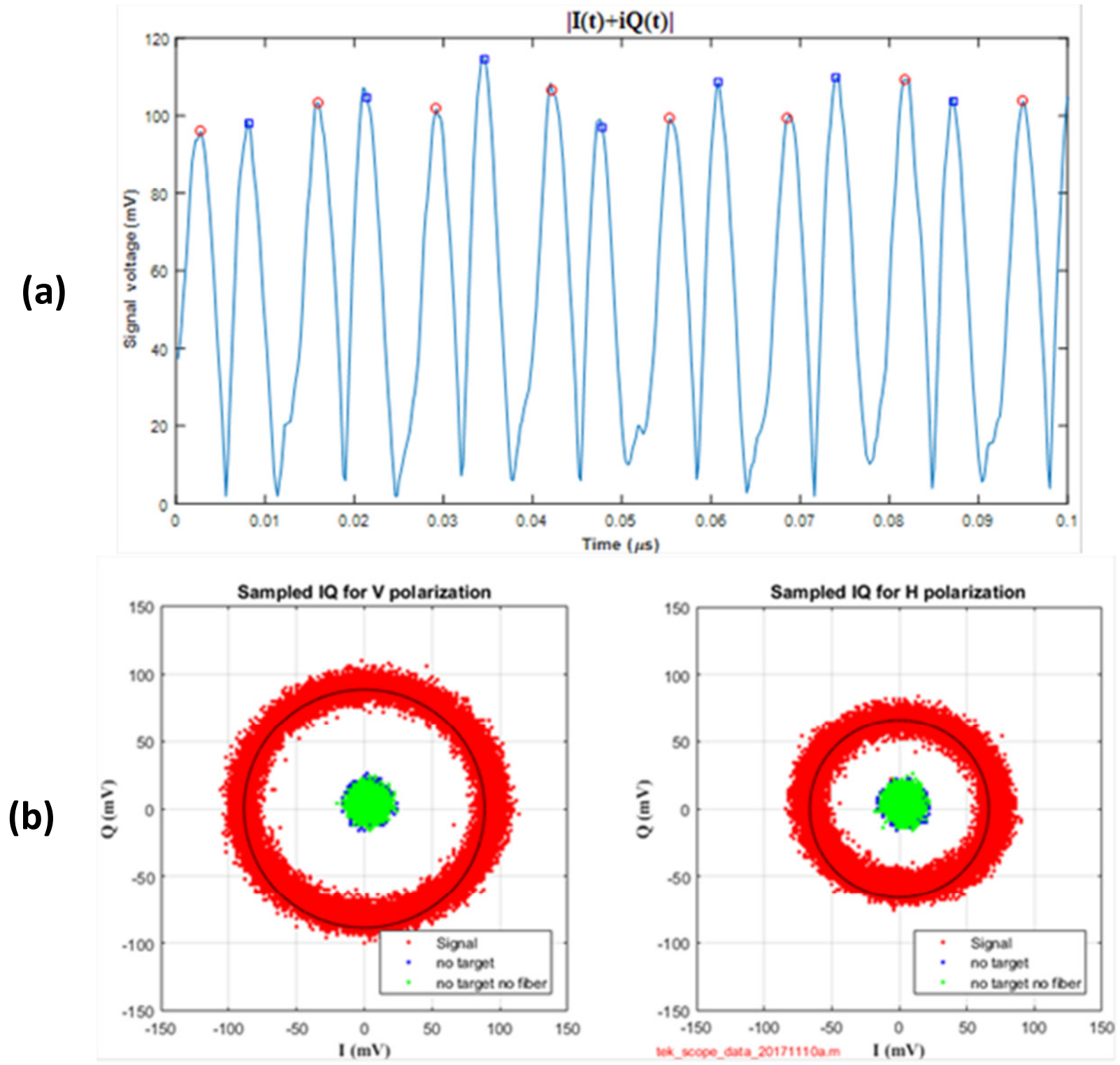

Fig. 4. Top: (a) Snap-shot in time of a typical amplitude of the complex reflectometer signal. The red circle and blue square markers denote the two orthogonal polarized pulse trains. Bottom: (b) The peak-sampled IQ plot for the two orthogonal polarizations for the cases with target, no target, and no target plus no fiber link. The black solid circle outline is a circle centered at the origin with a radius equal to the average amplitude of the IQ signal.

From the down-sampled signal, the two orthogonal polarizations V and H are separated into two sequences. This can be readily done since the V and H pulses are separated by ~ 5 ns while the pulse period is about 13 ns (~ 76 MHz PRR). The Stokes parameters can be readily computed using the two complex V and H signals.

To reduce the impact of noise, spectral filtering was performed on the complex signals to remove the DC as well as the high frequency components. The DC component is a result of the imperfect balanced detection described earlier. The filtered complex signals are then used to compute $S_a$, the time-average Stokes vector length. The data processing steps for this are summarized as follows:

- For each reference arm delay, a complex signal, $I(t) + iQ(t)$, is acquired and sampled.
- The two orthogonal polarizations, $V$ and $H$, are extracted from the complex signal at or near the pulse peak, yielding two sets of complex signals, $I_{Vn} + iQ_{Vn}$ and $I_{Hn} + iQ_{Hn}$.
- An FFT is performed on the two sampled polarization signal components to generate complex spectra $S_{VIQ}(f)$ and $S_{HIQ}(f)$ for the two polarization components.
- The complex spectrum pair is filtered to remove specific spectral components.
- An inverse FFT is performed on the filtered spectrum pair yielding $IQ_{Vf} = I_{Vf} + iQ_{Vf}$ and $IQ_{Hf} = I_{Hf} + iQ_{Hf}$.
- The Stokes parameters and time-averaged Stokes vector length are computed using [7]

$$S = \begin{bmatrix} S_0 \\ S_1 \\ S_2 \\ S_3 \end{bmatrix} = \begin{bmatrix} |IQ_{Hf}|^2 + |IQ_{Vf}|^2 \\ |IQ_{Hf}|^2 - |IQ_{Vf}|^2 \\ 2\,\mathrm{Re}\{IQ_{Hf} \times IQ_{Vf}{}^*\} \\ -2\,\mathrm{Im}\{IQ_{Hf} \times IQ_{Vf}{}^*\} \end{bmatrix},$$

and

$$S_a = \sqrt{\langle S_1\rangle_t^2 + \langle S_2\rangle_t^2 + \langle S_3\rangle_t^2}.$$

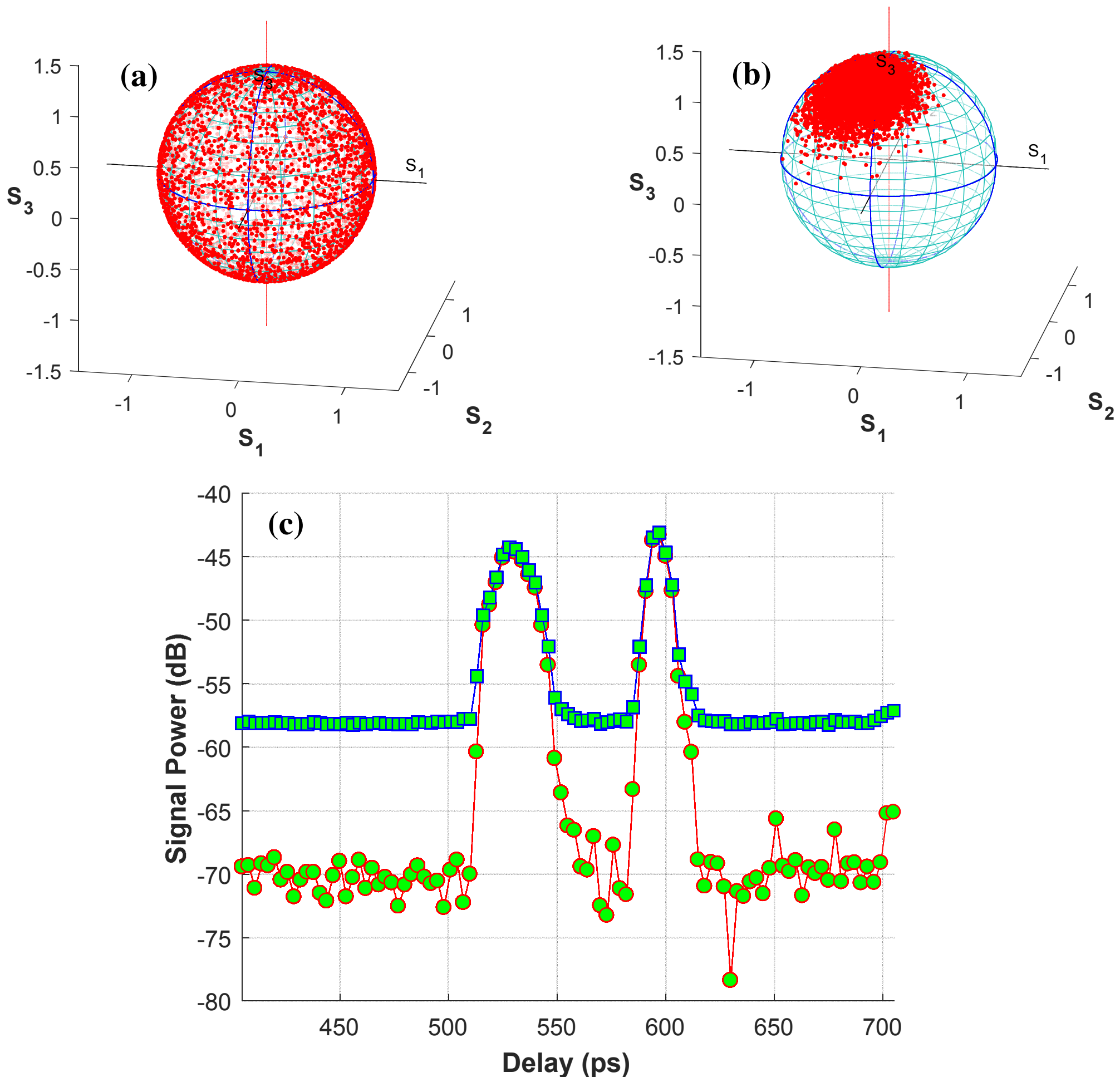

Fig. 5. (a) Points on the Poincare sphere show the Stokes vector when the interferometer is monitoring a location some distance away from the target; (b) the Stokes vectors when the interferometer is monitoring the reflection from a small fiber Bragg grating with 1% reflectivity; (c) Plot of the signal as a function of the reference arm time delay when scanning over a target which consists of two small Bragg Gratings separated by 8 mm. The top trace is the signal with spectral filtering described in the text. The bottom trace is the signal with spectral filtering and with averaging of the Stokes vectors.

Fig. 5a and 5b show plots of Stokes vectors calculated for the signal reflected from the target, and the signal when the reflectometer is tuned off target by detuning the PRR or the VDL. Fig. 5c shows the observed signal power $\langle S_0 \rangle_t$ and the averaged Stokes vector $S_a$ when the interferometer was tuned to observe a target consisting of a pair of fiber Bragg gratings. We see that the signal from the grating is barely affected by averaging, while the RBS is reduced by an order of magnitude.

## 4. Phase monitoring and correction

The complex $IQ$ signal from the coherent receiver will be affected by $\theta_L(t)$, the slow phase variation of the laser source. In the absence of other disturbances, the $IQ$ signal phase should vary as $\theta_{IQ}(t) = \theta_L(t) - \theta_L(t - \Delta t)$, where $\Delta t$ is the propagation time difference between the test arm and the reference arm.. In order to compensate for the phase variation, the phase $\phi_L(t)$can be experimentally determined using the auxiliary interferometer, and then canceled out in signal processing.

The output of the auxiliary MZI, which has a transit time difference δt, will be amplitude modulated by the time-varying laser phase fluctuations and is proportional to $I_{MZI} \propto \cos[\phi_{MZI}(t)]$ , where $\phi_{MZI}(t) = \phi_L(t) - \phi_L(t - \delta t)$ . The phase, $\phi_{MZI}(t)$, can be reconstructed from experimental data by first calculating the Hilbert transform of the modulated signal $I_{MZI}$, which will be approximately proportional to $\sin[\phi_{MZI}(t)]$. The phase difference is then given by $\phi_{MZI}(t) = \tan^{-1}\left[\frac{Hilbert(I_{MZI})}{I_{MZI}}\right]$. Since the arctangent function maps the phase angle to the range [-π, π ], the phase will need to be unwrapped.

Assuming that $\delta t$ is small enough, one can write $\phi_{MZI} = \delta t \frac{d\phi_L}{dt}$, which can then be used to extract the reconstructed phase of the laser $\phi_L(t)$ using $\phi_L(t) = \pm \frac{1}{\delta t} \int_0^t \phi_{MZI}(t')dt'$.

where the $\pm$ indicates the ambiguity in the sign of $\phi_L(t)$. If a coherent receiver is used to measure the output of the auxiliary MZI, we would be able to determine the laser phase unambiguously.

In order to see how well the phase of the complex signal from the reflectometer can be reconstructed, the phase $\theta_{IQ}(t) = angle(IQ)$ can be plotted verses the reconstructed phase $\phi_{IQ}(t) = \phi_L(t) - \phi_L(t - \Delta t)$. This plot is shown in Fig. 6 below for typical data. This figure illustrates the sign ambiguity and that the reconstructed phase is the same as the measured phase to within a constant phase shift.

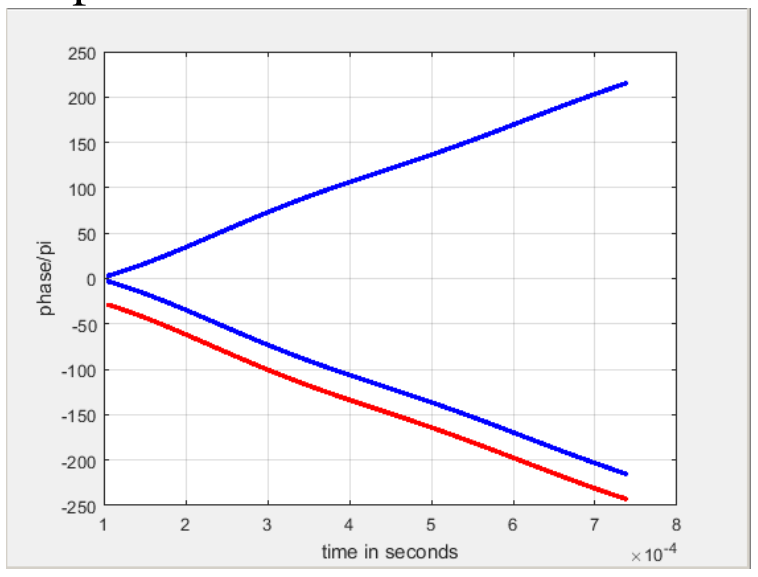

Fig. 6. Plots of the measured phase of the reflected signal from a target (red) $\theta_{IQ}(t)$ and $\pm\phi_{IQ}(t)$, the reconstructed phase from the auxiliary MZI (blue). One of the signs of $\phi_{IQ}(t)$ tracks the real phase.

Once we have determined $\phi_L(t)$, along with its correct sign, a phase correction can be applied to the complex $IQ$ signal for each of the polarizations. Fig. 7 shows the spectrum for a selected set of data samples for short periods of time, on the order of 100 microseconds, and the corresponding spectrum of $IQe^{-i\phi_{IQ}(t)}$ for three different data sets for one polarization.

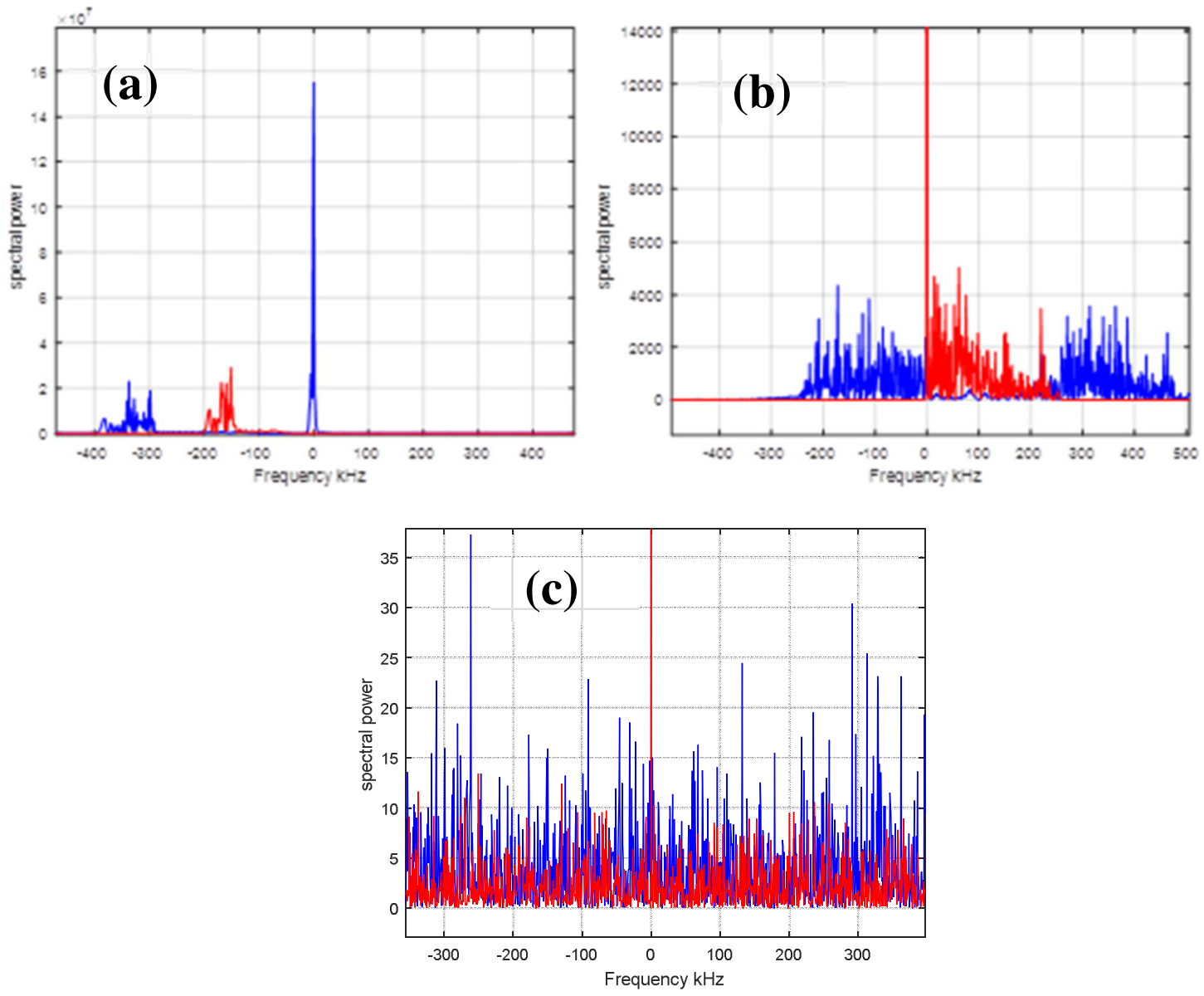

Fig. 7. Spectra of three different sample data sets, without the phase correction (red), and with the phase correction (blue). (a) Signal from target. (b) Fiber without target, and (c) without any fiber.

In order to determine the Rayleigh backscattering background, the target is disconnected from the test fiber link, and the same process of reconstructing the signal's phase from the auxiliary MZI data is repeated and used to correct the signal.

The signal is calculated by first removing all low frequency components within +/- 10 kHz of the center, which corresponds to the repetition rate of the mode locked laser. This eliminates the center spike shown in Fig. 2, and corrects for the small imbalance of the coherent receiver due to the finite common-mode rejection ratio of the balanced detectors. We perform an inverse Fourier transform, and apply the phase correction using the reconstructed phase from the auxiliary MZI. The corrected signal is again Fourier transformed. In the case of a reflected signal from a single target (Fig 7a), we see a dramatic reduction on the spectral width. This is does not occur for the case of fiber only without target (Fig 7b) or when there is no fiber (Fig 7c).

The spectral power of the corrected data within +/- 10 kHz of the center is then integrated. When the average SNR is calculated using the no-target data as the noise background, the resulting SNR is 43 dB, while when using the no-fiber background data the SNR is 69 dB. The reflectivity of the target was approximately 1%. This implies that our reflectometer should be able to detect a target with a reflectivity of -63 dB in the presence of Rayleigh backscattering background when the SNR=1.

In the case of no Rayleigh backscattering, targets should be detectable down to target reflectivities of -89dB. This performance is still several times worse than the theoretical shot noise limited performance estimated to be about -100 dB. This discrepancy is most likely due to losses in the detection optical components and insufficient power in the local oscillator. While this can be readily corrected, it will not improve discrimination against the Rayleigh backscattering noise. To further suppress RBS, either the strength of laser phase modulation must be increased, or the polarization averaging must be performed on a time scale longer than $T_{coh}$.

## 5. Conclusions and future work

In this work, we demonstrated a concept for a sensitive reflectometer with high spatial resolution for long fiber optic links. A mode locked laser with a constant pulse repetition rate was used. This resulted in a significant increase in the noise due to RBS. Two different techniques for reducing the RBS noise were demonstrated. The Stokes vector averaging results in an order of magnitude improvement. Monitoring of the laser phase with an auxiliary interferometer, and using the extracted phase to correct the signal, results in increased sensitivity and reduction of RBS. This approach could potentially be improved further with enhanced laser phase bandwidth, which could be readily accomplished with an electro-optic phase modulator.

## References

1. D. Derickson, Fiber optic test and measurement, New Jersey: Prentice Hall, 1998.
2. M. Legré, R. Thew, H. Zbinden and N. Gisin, "High resolution optical time domain reflectometer based on 1.55μm up-conversion photon-counting module," Opt. Express, **15**(13), 8237–8242 (2007).
3. "Optical Backscatter Reflectometer™ 4600," Luna Innovations, Inc., 2015. [Online]. Available: http://lunainc.com/obr4600.
4. A. Stove, "Linear FMCW radar techniques," IEE Proceedings-F, **139**(5), 343-350 (1992).
5. J. Minar, H. de Riedmatten, C. Simon, H. Zbinden and N. Gisin, "Phase-noise measurements in long-fiber interferometers for quantum-repeater applications," PHYSICAL REVIEW A, vol. 77, pp. 052325-1, 2008.
6. T. Ahn, J. Lee and D. Kim, "Suppression of nonlinear frequency sweep in an optical frequency-domain reflectometer by use of Hilbert transformation," Applied Optics, **44**(35), 7630-7364 (2005).
7. M. Born and E. Wolf, *Principles of Optics* (Cambridge University Press, 1999)